# Main Manuscript for

# Prospects for Detecting Signs of Life on Exoplanets in the JWST Era

Sara Seager[1,2,3,*], Luis Welbanks[4], Lucas Ellerbroek[5], William Bains[6], Janusz J. Petkowski[7,8]

[1] Department of Earth, Atmospheric and Planetary Sciences, Massachusetts Institute of Technology, Cambridge, MA, USA
[2] Department of Physics, Massachusetts Institute of Technology, Cambridge, MA, USA
[3] Department of Aeronautical and Astronautical Engineering, Massachusetts Institute of Technology, Cambridge, MA, USA
[4] School of Earth and Space Exploration, Arizona State University, Tempe, AZ, USA
[5] Department of Astrophysics / IMAPP, Radboud University, PO Box 9010, 6500 GL, Nijmegen, The Netherlands
[6] School of Physics and Astronomy, Cardiff University, 4 The Parade, Cardiff CF24 3AA, UK
[7] JJ Scientific, Warsaw, Mazowieckie, Poland
[8] Faculty of Environmental Engineering, Wroclaw University of Science and Technology, 50-370 Wroclaw, Poland

*Corresponding author: Sara Seager

**Email:** seager@mit.edu
**Preprint Servers:** N/A.
**Author Contributions:** Writing—original draft preparation (S.S). Writing—review and editing (L.W., L.E., W.B., J.J.P.). All authors have reviewed the manuscript and agree with the results.
**Competing Interest Statement:** The authors declare no conflict of interest.
**Classification:** Physical Sciences; Earth, Atmospheric, and Planetary Sciences
**This PDF file includes:** Main Text; Figures 1 to 5; Table 1.

## Abstract

The search for signs of life in the Universe has entered a new phase with the advent of the James Webb Space Telescope (JWST). Detecting biosignature gases via exoplanet atmosphere transmission spectroscopy is in principle within JWST's reach. We reflect on JWST's early results in the context of the potential search for biological activity on exoplanets. The results confront us with a complex reality. Established inverse methods to interpret observed spectra—already known to be highly averaged representations of intricate 3D atmospheric processes—can lead to disparate interpretations even with JWST's quality of data. Characterizing rocky or sub-Neptune-size exoplanets with JWST is an intricate task, and moves us away from the notion of finding a definitive "silver bullet" biosignature gas. Indeed, JWST results necessitate us to allow 'parallel interpretations' that will perhaps not be resolved until the next generation of observatories. Nonetheless, with a handful of habitable-zone planet atmospheres accessible given the anticipated noise floor, JWST may continue to contribute to this journey by designating a planet as biosignature gas candidate. To do this we will need to sufficiently refine our inverse methods and physical models for confidently quantifying specific gas abundances and constraining the atmosphere context. Looking ahead, future telescopes and innovative observational strategies will be essential for the reliable detection of biosignature gases.

## Main Text

### 1. Introduction

The quest to understand life beyond Earth is one as old as humanity itself. Since the earliest days of modern astronomy, the presence of oxygen ($O_2$) in Earth's atmosphere was recognized as due to, and hence to be a signature of, life (1). Oxygen appeared in Earth's atmosphere at least 2.7 billion years ago (2) when primitive cyanobacteria evolved to perform photosynthesis, converting carbon dioxide ($CO_2$) and water ($H_2O$) into carbohydrates and oxygen, using sunlight. This process changed the chemical composition of Earth's atmosphere by introducing $O_2$.

From an exoplanetary perspective, Earth's 20% atmospheric oxygen is anomalous due to oxygen's high reactivity, sustained by oxygenic photosynthesis. An extraterrestrial civilization with advanced telescopic technologies observing Earth, could interpret $O_2$'s high concentration as a strong indicator of life.

Oxygen is the paradigmatic example of a "biosignature gas", a gas that is produced by life and accumulates to high enough atmospheric abundances to be detected with remote telescopes. Yet despite evidence pointing to the very early origin of oxygenic photosynthesis 3.5 Gya (3), $O_2$ took billions of years to accumulate to its present-day values of 20% even whilst $O_2$-producing life was present (4). Therefore, during much of the time when we are confident that Earth was inhabited, $O_2$ was at most a trace component of the atmosphere.

Molecules other than $O_2$ should therefore be considered as potential biosignature gases. Earth's biosphere generates thousands of different volatile molecules for various reasons—as waste products from exploiting chemical potential energy gradients, for signaling or defense against predators, to name a few—which could be mirrored or replaced by different compounds on other worlds. Indeed, aside from the noble gases, every gas in Earth's atmosphere to part-per-trillion levels is produced by biological activity (5), although most gases have a primary source other than life. Extensive exploration of different gases, not just those produced in significant quantities by life on Earth, is needed.

We now have our first real opportunity to search for exoplanet atmosphere biosignature gases with the recently operational James Webb Space Telescope (JWST; (6))—a search that is becoming an expanding area of research within Astrobiology. We aim to assess the present opportunities and challenges to using the JWST to search for exoplanet atmosphere signs of life. We begin with a review of JWST's enabling photometric precision as it relates to target planet numbers (Section 2) and continue with prospects and pitfalls via current exoplanet atmosphere interpretation methods (Section 3). We directly discuss the prospects for JWST to detect biosignature gases (Section 4), and conclude with future needs including innovative methods and new telescopes (Section 5).

## 2. JWST Transmission Spectroscopy Precision and Habitable-Zone Targets

To understand the prospect of detecting biosignature gases with JWST we summarize a defining JWST observation related to its data precision and noise floor, and estimate what this noise floor translates into for numbers of targets suitable for biosignature gas searches. JWST uses transmission spectroscopy (7, 8), the dimming of a star as a planet transits across its face, allowing us to analyze the starlight filtered through the planet's atmosphere (Figure 1). This method reveals variations in atmospheric composition across different wavelengths due to selective absorption by atmospheric gases. Other JWST atmosphere observation methods do not reach habitable-zone temperate exoplanets[1]. Such a planet's low infrared emission will be overshadowed by radiation from their host stars, making them unsuitable for secondary eclipse spectroscopy, and their small planet-star separations makes them unsuitable for direct imaging.

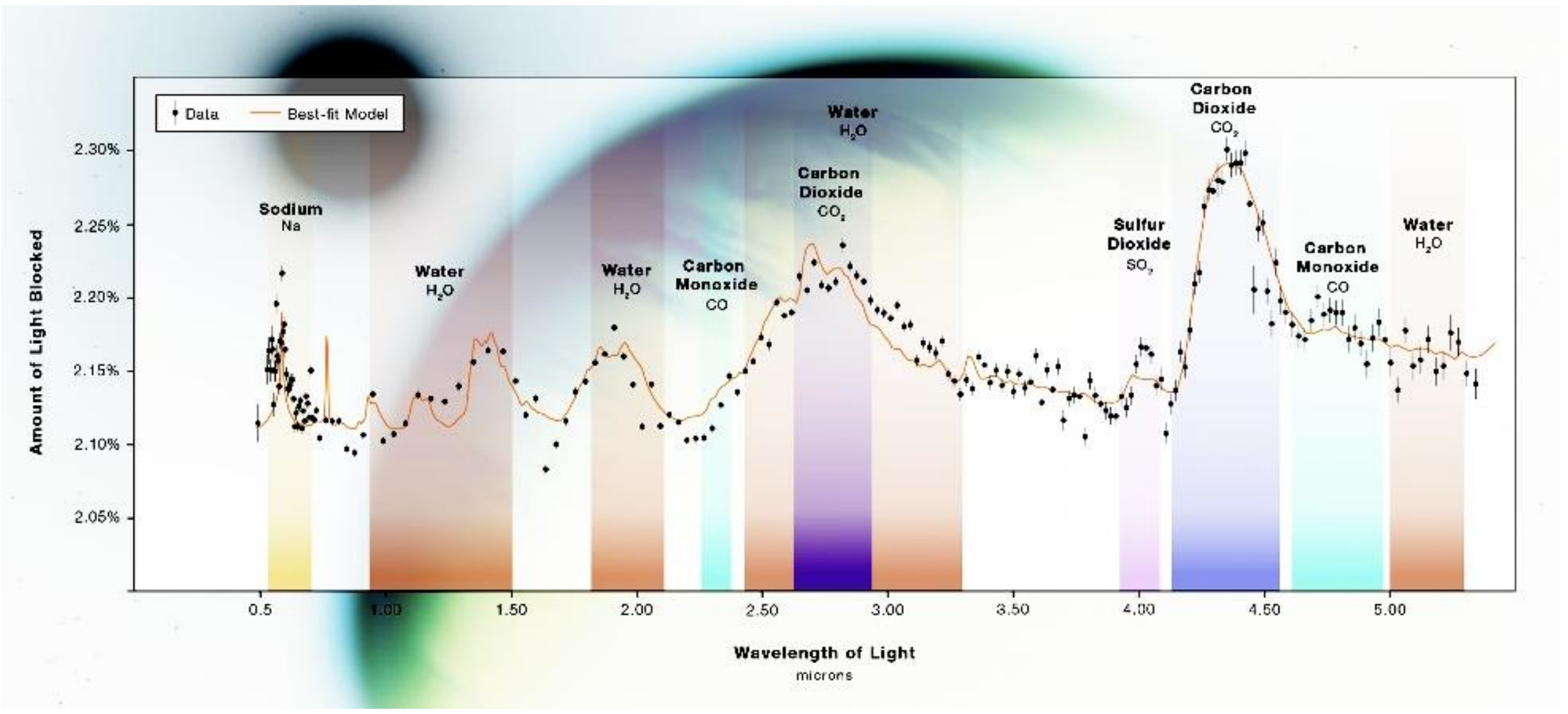

Figure 1. Wasp 39b transmission spectrum. The planet size (shown as amount of light blocked) changes with wavelengths, increasing where individual gases are strongly absorbing. Figure credit: NASA, ESA, CSA, STScI.

Among JWST's many notable observations, we highlight an early, striking demonstration of JWST's unprecedented precision, leading to the unequivocal, robust detection of carbon dioxide ($CO_2$) in the atmosphere of the "hot Saturn" exoplanet WASP-39b ($M_p$ =

[1] The habitable zone is the region around a star in which a planet may maintain liquid water on its surface, based on its surface temperature for a planet atmosphere heated by the radiation from the host star.

0.28 $M_J$, $R_p$ = 1.27 $R_J$, P = 4.1 days, $T_{eq}$ ~1100 K; Figure 1; (9)). The detection of $CO_2$ is striking because $CO_2$'s spectral band is clearly visible "by eye" in the data. Despite not being the dominant carbon-bearing species in WASP-39b's hydrogen-dominated atmosphere, $CO_2$ strongly absorbs infrared light, creating a distinct spectral feature. Sulfur dioxide ($SO_2$), also detected in WASP-39b's atmosphere, arises from photochemical reactions rather than predicted chemical equilibrium (10). This highlights the complexities of exoplanet atmospheres, and marks a shift towards a new subfield of astrochemistry. We emphasize that WASP-39b is a giant exoplanet too hot for life beneath its massive hydrogen-helium envelope.

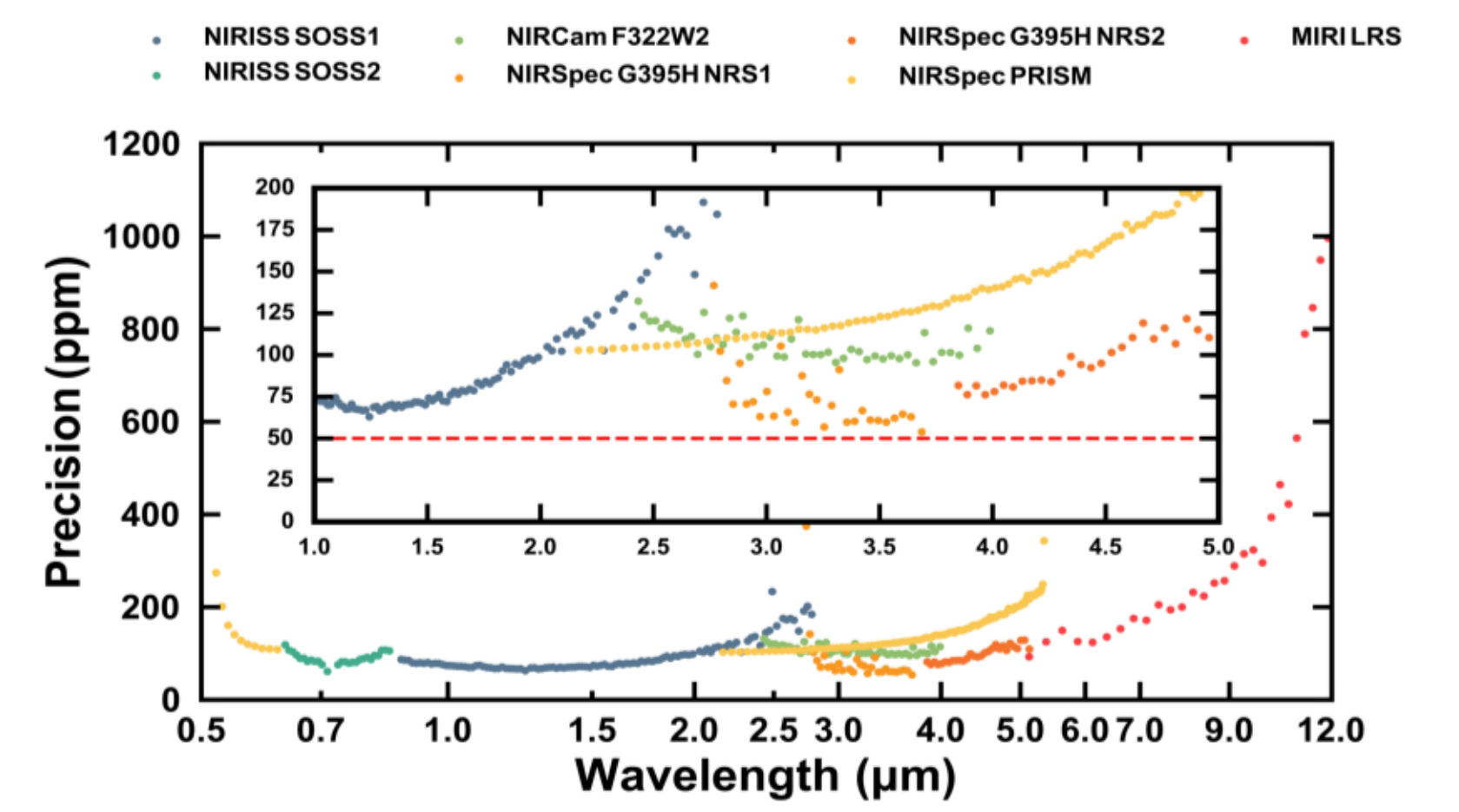

Figure 2. Spectroscopic photometric precision of WASP-39b. The precision (y-axis) as a function of wavelength (x-axis) for different JWST instruments (colors, as noted in the legend). As an Early Release Science Program target, WASP-39b was observed with each JWST instrument, NIRISS SOSS (0.6—2.8 μm), NIRSpec G395H (2.7—5.2 μm), NIRSpec PRISM (0.5—5.5 μm), NIRCam F322W2 (2.0—4.0 μm), and MIRI-LRS (5.0—12.0 μm). While the precision in the measured transit depth is chromatic and dependent on the stellar spectrum, the spectrum of WASP-39b shows precisions below 200 ppm between 1.0—5.0 μm. Not shown is that the spectroscopic light curve data precision is close to the photon-limited noise (9). Data from (56) and Welbanks (priv. comm).

With one transit for the relatively bright star WASP-39, JWST has achieved 50 ppm precision (Figure 2), demonstrating performance of the telescope at the spectroscopic light curve close to the photon-limited noise (9). The 50 ppm value is similar to that achieved by other observational programs for sub Neptunes and super-Earths (11–13). JWST's true noise floor can only be assessed with several combined transits, to disentangle photon noise from actual instrumental limitations. Although this assessment is ongoing, we adopt a 30 ppm noise floor value for the purposes of this review.

By noise floor, we mean the minimum level of unwanted signal arising from a combination of factors related to the instrument and observational environment (e.g., thermal fluctuations, mechanical vibrations, characteristics of the infrared detectors, and unknown sources) (14,15). This variety of contributing factors makes the noise floor best determined empirically.

We can qualitatively investigate what kind, and how many, exoplanets in their host star's habitable zone are accessible under the 30 ppm noise floor. We estimate the transmission spectroscopy signal (*TS*), by taking the area of a three scale height (*H*) tall atmosphere annulus compared to the area of a homogeneous background star.

$$TS \approx \frac{6HR_p}{R_*^2} \quad \text{(Equation 1)}$$

Here $R_p$ and $R_*$ are the planet and star radius respectively. $H$ is defined as $H = kT/mg$, where $k$ is Boltzmann's constant, $T$ is temperature, $m$ is mean molecular mass, and $g$ is surface gravity.

We can immediately rule out the JWST transmission spectroscopy study of the atmosphere of an Earth-like planet in an Earth-like orbit around a Sun-like stars from Equation (1). Namely, Earth-like planets' small sizes ($R_p$ ~ 6400 km) and thin atmospheres ($H$ = 8 km) are impractical for transmission spectroscopy against the backdrop of a Sun-sized star ($R_*$ ~ 700,000 km). Here $TS$ ~ 1 ppm—far lower than the adopted 30 ppm noise floor.

The observational favorability of M dwarf stars can also be seen from Equation (1). Since M dwarf stars are half to one tenth of the size of our Sun, the $TS$ will have signals 4 to 100 times larger than Sun-sized star hosts.

We can further use the approximate $TS$ feasibility criteria to estimate the number of habitable-zone exoplanets with accessible atmospheres. Considering the list of currently observed or planned exoplanet targets from the JWST Cycles 1 through 3 from the Transiting Exoplanets List of Space Telescope Spectroscopy Catalog (TrExoLiSTS) (14) and estimating the surface temperature by the equilibrium temperature ($T_{\mathrm{eq}}$) with a cutoff of 300 K for habitable-zone planets, we find only a few Earth-size planets ($R_p \leq 1.5\ R_{Earth}$), namely some of the Trappist-1 exoplanets and LP 791-18d. (Here $T_{\mathrm{eq}}$ is the effective temperature attained by an isothermal planet atmosphere after it has reached complete equilibrium with the radiation from its parent star.

Sub Neptunes are of interest in the search for life not necessarily because they are expected to be habitable, but because their likely $H_2$-dominated atmospheres with scale heights 14 times larger than $N_2$-dominated atmospheres make them more favorable for atmospheric observations. Some sub Neptunes are hypothesized to have liquid water under the right conditions (15, 16). Regardless of the presence of water oceans, some sub Neptunes may have water clouds, offering the potential for an aerial biosphere (17).

As many as half a dozen sub-Neptune-sized exoplanets ($1.5\ R_{Earth} \leq R_p \leq 3\ R_{Earth}$) meet our $TS$ feasibility criteria, even with a conservative $T_{eq}$ cutoff of 373 K (100°C). However, these planets are also likely too hot for life due to greenhouse effects from $H_2$ collision-induced absorption (18). Observational biases from K2's and TESS's viewing segments have favored shorter orbital periods, but ongoing TESS observations should uncover sub-Neptunes with longer periods and therefore cooler temperatures.

Although M dwarf stars are the only accessible temperate planet-hosting targets for JWST, M dwarf stars present a challenge. Their stellar magnetic activity, higher than for Solar-type stars, manifests as star spots, faculae, and flares that contaminate the spectra (19). Star spots can mimic transit transmission spectra by the star's inhomogeneity (Figure 3). In the case of Trappist-1, the measured star contamination dominates the signal (12). The community is intensively pursuing many ideas to mitigate or remove contamination signals (19). Some stochastic variability might never be adequately modeled and may just have to be accepted as part of the noise floor.

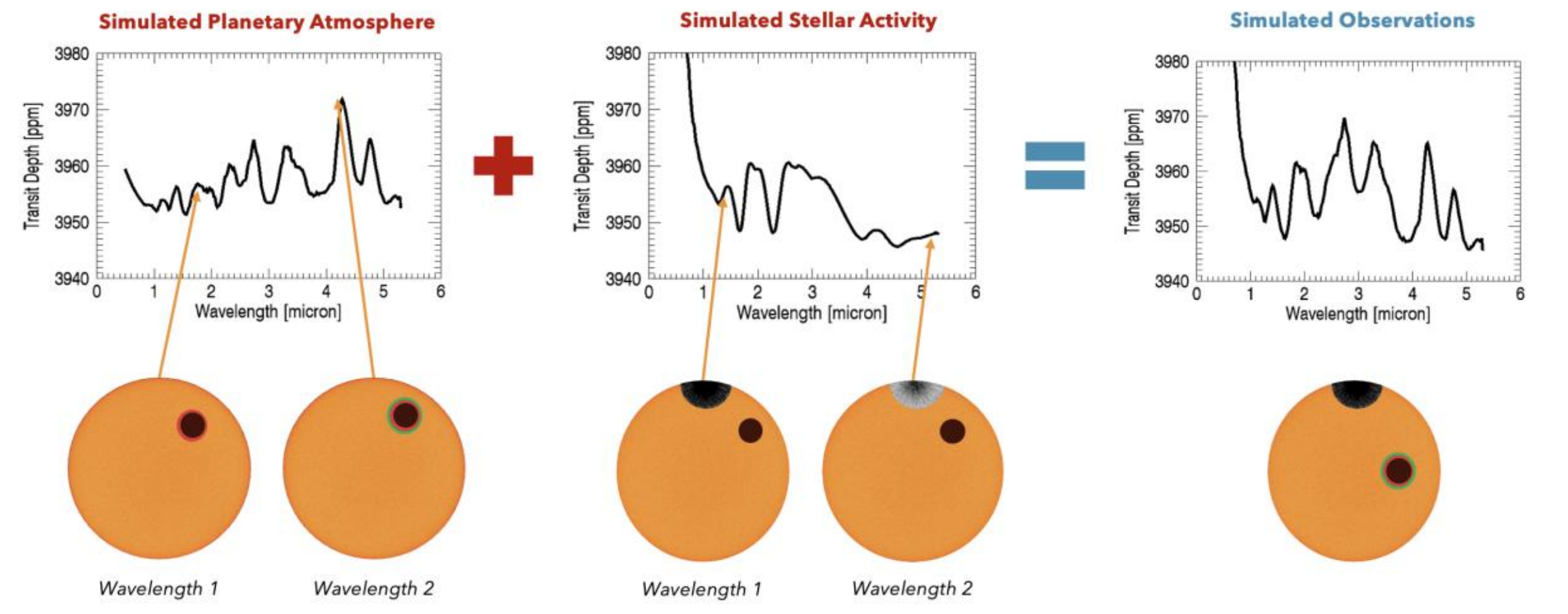

Figure 3. Star spot contamination in transit transmission spectra. Left: Transit depth varies with wavelength as a planet (black circle) transits a star (orange disk) with an atmosphere (colored annuli), due to wavelength-dependent atmospheric absorption. Middle: A star spot (black/gray semi-circle) alters the star's apparent size with wavelength, affecting the transit depth even for a planet without an atmosphere. Right: The combined spectrum shows contributions from both the planet's atmosphere and the star spot, complicating interpretation. Credit: Seager and Shapiro.

## 3. From Spectroscopy Measurements to Planetary Characterization

To understand the JWST's prospects for identifying biosignature gases, we begin by examining ongoing planet characterization. We unpack the process of translating observations into estimates of exoplanet properties, beginning with a general description, then proceeding to a case study on sub-Neptunes, and concluding with the relevance for biosignature gases.

It may seem a stretch to use spectra to ascertain planetary properties (atmosphere abundances, surface and interior bulk composition, habitability and presence of life, and more). After all, observed exoplanet spectra represent a highly averaged signal of complex 3D physical and chemical atmospheric processes, reduced to relative changes in the observed wavelength-dependence of the combined star and planet light as a point source. We live with a family of potential solutions for planet properties that fit the current imperfect and incomplete data. To quantify the range of planet properties, the community resorts to atmosphere retrieval (20), an umbrella term for a subset of inverse methods. This process determines the underlying atmospheric parameters that best match the observations for a given model (see e.g., the review by (21)). The outcome is not only an assessment of how well a specific model fits, but also a statistical measure of the likely values for the model's parameters, such as probable abundances of atmospheric gases.

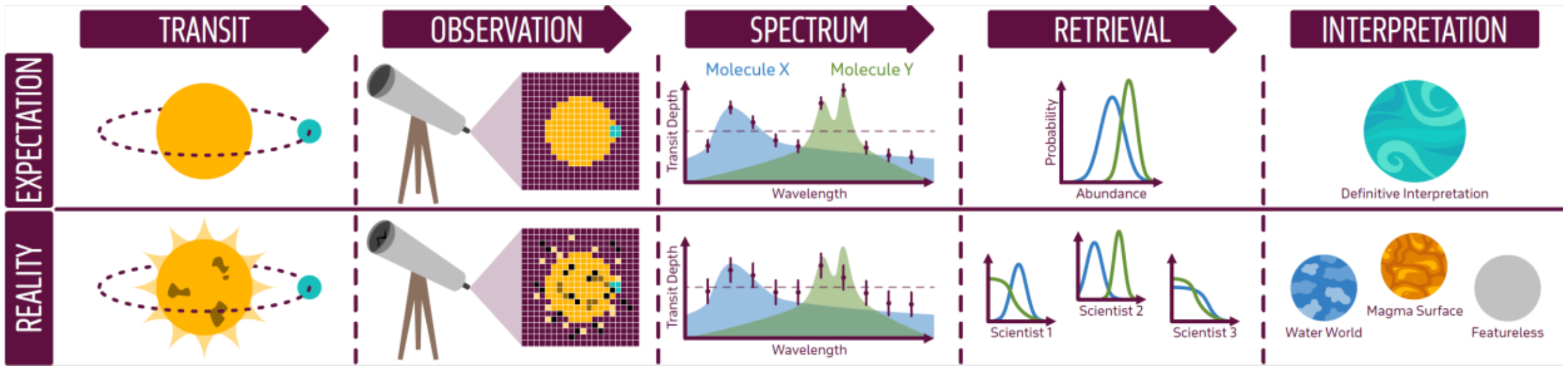

Figure 4. Illustrative schematic of the atmospheric data analysis and interpretation process, from observations through interpretation of an exoplanet's nature. The pathway involves four main steps: [1] taking observations; [2] processing the data from pixels to a spectrum; [3] interpreting the resulting spectrum with model fits to infer planetary properties such as gas abundances; and [4] drawing conclusions about the planet properties. The top row is inspired by WASP-107b (35) a relatively straightforward case, and the bottom row is motivated by TOI 270d ((29)), a disputed planetary nature. All panels are illustrative only. See text for details. Credit: Seager, Welbanks, Tilke.

Inferring planet properties from spectroscopic measurements is not straightforward as the process involves a number of steps with subjective choices (Figure 4). Step 1 is the telescope observations (i.e., raw pixel data). Step 2 is converting the observations into a planetary spectrum via a data analysis pipeline, which always involves a number of assumptions. Step 3 is producing the family of models that fit the data—inferring the underlying planet atmosphere properties (i.e., temperature, gas abundances) from the data. Step 3 is challenging due to the large number of free parameters in most models compared to the number of data points, and the ensuing degeneracies between these parameters. These challenges cause atmospheric retrieval to be highly sensitive to the choice of model parameterization and model assumptions (see e.g., (22)), and in some cases to small changes in the data and the structure of the data uncertainties. This can result in inferred planet properties that are ultimately incorrect—and rejected by standard frequentist metrics such as chi-squared statistics and p-value hypothesis testing. Moreover, most atmospheric retrievals rely on the assumption of Gaussian uncorrelated noise, whereas the data retrieved on has known correlated systematic noise. Step 4—planetary interpretation—is not only the end goal but also the motivation for studying exoplanet atmospheres. Step 4 includes identifying the planet archetype (e.g. “water world” or “lava planet”), habitability, presence of biosignature gases, and more. Unfortunately, Step 4 is the most subjective in the entire process of exoplanetary characterization, and is still evolving. In some cases, the uncertainties arising from Steps 1-3 are so significant that they cannot rule out fundamentally different planet archetypes. In other situations, the planet’s atmosphere is simply unrevealing of its nature**.** Unveiling the 'truth' of a planet with incomplete and imperfect data and models is the ultimate test of interpolation and extrapolation, yet a challenge we must embrace to characterize exoplanets.

To illustrate the challenges faced by the atmospheric data analysis and interpretation framework let us turn to examples of sub Neptunes. Sub Neptunes defy straightforward classification because their average densities match a variety of bulk interiors. They could be: water worlds with 50% or more water by mass with liquid water (or supercritical water) ocean; scaled-down versions of Neptune with a thick $H_2$-He envelope overlaying a layer of hot dense water plasma over a rocky core; they could be predominantly composed of silicates, but with a mixed hydrogen envelope possibly overlying a magma ocean (e.g. (23)); or even scenarios we have yet to consider. The community has aimed to use atmosphere measurements to sort through the possibilities for sub Neptune archetypes.

A classic example of the challenges in every step in the data-to-characterization process is the case of K2-18b ($M_p$ = 8.6 $M_E$, $R_p$ = 2.6 $R_E$, P = 32.94 d, $T_{eq}$ ~ 260 K) (24, 25). Hubble Space Telescope (HST) observations led to a detection of atmospheric water vapor in the atmosphere of K2-18b (24, 26). Yet, later observations with the JWST found no water vapor, and attributed the same atmospheric spectral feature to $CH_4$ (16). Furthermore, the non-detection of $NH_3$ in K2-18b's atmosphere was presented as a case for a Hycean world (a planet with a liquid water ocean covered with a $H_2$-dominated atmosphere), since $NH_3$ is highly soluble in liquid water thus explaining its absence (16). This conclusion was later disputed, as nitrogen compounds are also highly soluble in magma, and therefore the lack of $NH_3$ could signify presence of a magma ocean, not a water ocean (27). Therefore, it is important to recognize that the challenges in Step 2 to 3 have cascading consequences all the way to the fourth and final step—the end goal of planet characterization.

A second example is the sub Neptune exoplanet TOI-270d ($M_p$ = 4.8 $M_E$, $R_p$ = 2.2 $R_E$, P = 11.38 days, $T_{eq}$ ~ 350 K) (28). The same observed pixels from JWST (Step 1) were processed by independent teams (Step 2). Both teams agree that the signal in the data is robust, and that the spectral features are attributed to $CH_4$ and $H_2O$. The inferred abundances $CH_4$ and $H_2O$, however, were different between the teams (Step 3). These differences, though small (1 to $2\sigma$) led to inconsistent interpretations of the planetary interior archetype (Step 4). One interpretation is a water-dominated planet with a hot surface water ocean (29), another a massive, mixed atmosphere of $H_2$, $H_2O$, and $CH_4$, overlaying a rocky massive interior (30).

The TOI-270d observations are also a cautionary tale for the JWST future of biosignature gas detection: the identification of both $CH_4$ and $H_2O$ still led to conflicting interpretations of planet archetype. Methane has been championed as a promising biosignature gas for JWST rocky planets—but $CH_4$'s biosignature status can only be inferred from the presence of other gases, such as the simultaneous presence of $CO_2$ in an atmosphere but little CO (31). Presently, low SNR data yield differences in quantified gas abundances of a factor ten to 100 or higher, so obtaining the necessary abundance constraints, key for establishing the $CH_4$ as a biosignature gas, may not be possible. We do, however, echo the argument in favor of $CH_4$ as a biosignature gas contender on rocky planets without $H_2$-dominated envelopes, due to its ease of detectability at JWST wavelengths (32) and its difficulty being produced in high quantities by known abiotic sources (for an oxidized mantle).

Further relevant to biosignature gases are that biosignature gases almost certainly produce weak signals, as they would originate from thin atmospheres, and likely have low abundances. In cases of low-SNR data with weak or statistically insignificant features, retrieval techniques are often employed to claim detections. A 'detection' is sometimes quantified by performing a model comparison between a reference model including the specific parameter being investigated (e.g., a given biosignature gas) and a model excluding the parameter of interest – as a pair of nested models. The model used as reference to claim a detection may be unphysical, may not fit the data appropriately, and may have other parameters with degeneracies relative to the parameter of interest. To recap, the oversensitivity to small changes in the data, the potential trap of comparing two incorrect models, and an opaque definition for the term ‘detection’ mean that using atmospheric retrieval can cause erroneous reports of detection of biosignature gases (33).

Let us step back and ask whether there are any straightforward cases of observations leading to definitive interpretation. Indeed, the case for the warm Neptune WASP-107b ($M_p$ = 0.096 $M_J$, Rp = 0.96 $R_J$, P = 5.72 days, $T_{eq}$ ~ 750 K) (34) presents such an example. Using our framework in Figure 4, WASP-107b was observed using JWST’s NIRCam (35) and NIRSpec (36) instruments (Step 1). Data were processed using different independent analysis pipelines (Step 2) and model fits via atmospheric retrieval agreed upon the assessed $CH_4$’s abundance. The inferred chemical abundances and temperature structure (Step 3), even with the small differences present in Step 2 independent analysis, indicate that the planet’s atmosphere is in a state of chemical disequilibrium, influenced by photochemistry and, likely, tidal heating (Step 4).

WASP-107b exemplifies a straightforward case because it excels in three key criteria critical to robust planet characterization: detection, attribution, and interpretation (see Section 4 for further details). The planet’s high SNR data ensures a robust detection of spectral features (Key Criterion 1: detection). Multiple spectral features of $CH_4$ were confidently attributed to the correct chemical species using different instruments and independent pipelines, which strengthens confidence in the association between the spectral features and the gases responsible for causing them (Key Criterion 2: attribution). Finally, the derived planetary properties — such as chemical abundances, temperature structure, and state of chemical disequilibrium — are consistent across independent analyses, allowing for a confident interpretation of the planet’s atmospheric state (Key Criterion 3: interpretation). These three criteria are essential for any atmospheric investigations, particularly those aimed at identifying biosignature gases, where precision in each of the three areas will be paramount.

Future exoplanet observations may not be as clear-cut as the WASP-107b example presented above. Even in this seemingly straightforward case with strong SNR across broad bands leading to high confidence in the presence of $CH_4$, and hence of chemical disequilibrium in the atmosphere, the results do not identify the driver of that disequilibrium. The best path forward is one of cautious optimism, acknowledging the limitations of our methods and data while striving to maximize the insights they offer. Regarding model interpretation, the complexity of the problem often exceeds the fidelity of even JWST data. Prioritizing high SNR features before delving into parameter inference will be essential for reliable interpretations. Simultaneously, advancing models and establishing physical “guardrails” are crucial steps to mitigate the subjective nature of the interpretation exercise. Perhaps for any “biosignature gas candidates”, we may hope for a situation similar to WASP-107b: an atmosphere highly out of chemical equilibrium (not due to UV-driven photochemistry), recognizable as such even under varying model assumptions.

## 4. Biosignature Gas Prospects with JWST

There are three Key Criteria for definitive exoplanet findings, that are also relevant for biosignature gases.

1. **Detection: Is the signal robust?**
   - For biosignature gases, the robustness of an unambiguous signal detection is highly relevant as any candidate signals are almost certainly going to be weak due to tiny atmospheres on rocky exoplanets and low anticipated atmospheric abundances.
2. **Attribution: Are the spectral features correctly attributed to the appropriate gas(es)?**
   - For biosignature gases, this involves identifying a distinctive spectral signature that stands out from dominant background atmospheric gases and that is unequivocally attributable to the candidate gas.
3. **Interpretation**: **How reliable are the derived planetary properties?**
   - For biosignature gases, this includes ensuring that the detected signal is not confounded by false positives (Table S2) and corresponds to plausible production rates.

In this Section we review current thinking on false positives and biosignature gas production rates (Section 4.1). We next present the growing list of biosignature gases, considering false positives and production rates (Section 4.2). We conclude with a review of the tentative claim of a JWST biosignature gas detection, as confronted against the above three Key Criteria (Section 4.3).

### 4.1 A Framework for Evaluating Potential of Biosignature Gases Detection

Ahead of observations, the community can make a list of promising biosignature gases by systematically assessing the potential of thousands of gases known to exist ((5); Figure 5). The process begins by determining whether a gas of interest has prominent and distinctive spectral features that stand out from anticipated background atmospheric gases (related to ease of detection and attribution). This is followed by considering potential false positives and, primarily, photochemical stability (both related to interpretation).

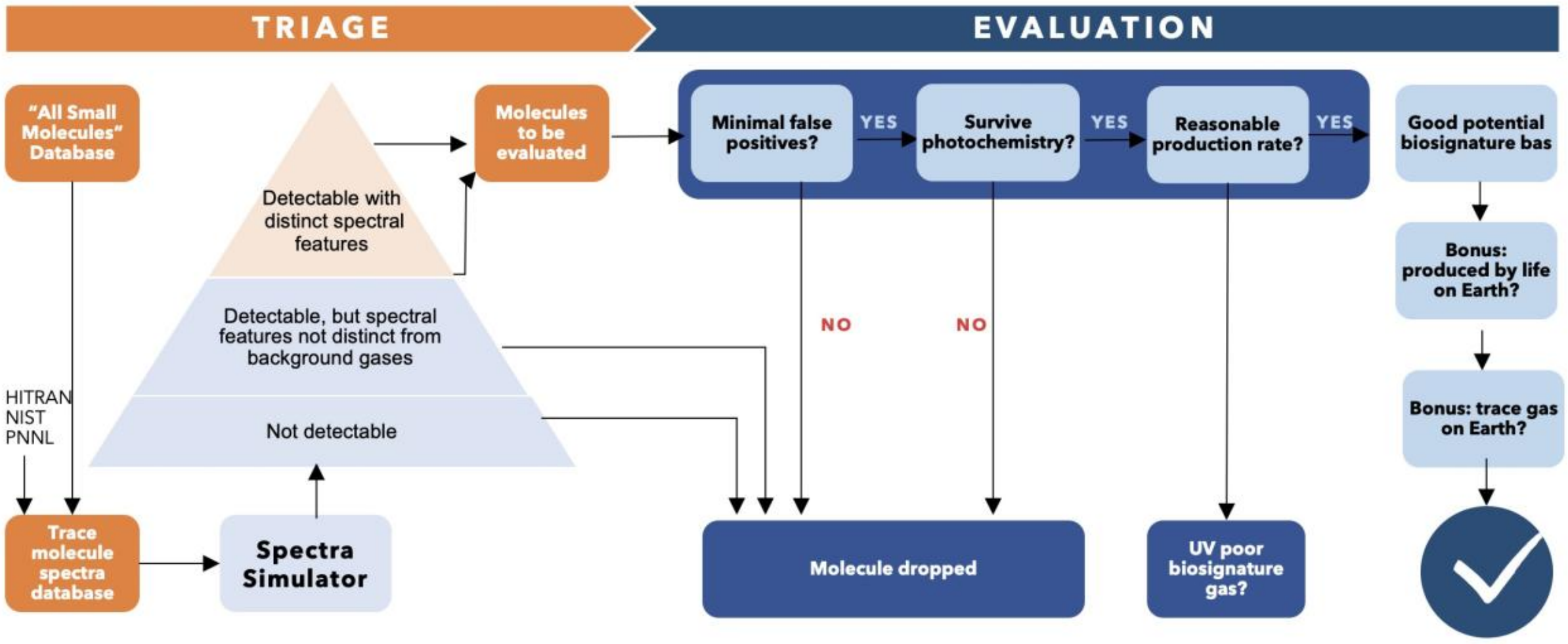

Figure 5. Schematic of a biosignature gas evaluation framework for thousands of gases. The first "triage" step identifies gases with prominent, distinctive spectral features. The second "evaluation" step focuses on photochemistry, assessing a molecule's atmospheric survival and whether biological production rates, compared to Earth's biofluxes, are plausible. Credit: Seager and Zhan.

<u>Biosignature Gas False Positives</u>

Fundamental to the interpretation of biosignature gases is the exclusion of false positives, that is gases that can be produced by abiotic processes as well as by life. Carbon dioxide ($CO_2$) is an obvious example, because although ubiquitously produced by life on Earth, $CO_2$ is also a significant background atmosphere gas, produced by volcanic processes and atmospheric photochemistry, and is therefore a dominant carbon species in planetary atmospheres. Molecular oxygen ($O_2$), despite being a favored biosignature gas, may have an abiotic source from evaporating oceans, originating from a runaway greenhouse effect (37), especially for a planet with close-to-host-star orbit or low mantle FeO and $H_2O$ inventories (38). For a review of $O_2$ and its false positives, see (39). One may estimate a potential biosignature gas's false positive propensity by a thermodynamic assessment of one or a pair of gases due to volcanic activity. (See the $CH_4$ vs. $CO_2$ vs. CO example in Section 3)

One concept of robust identification against false positives is that of chemical disequilibrium—the simultaneous presence of reduced and oxidized species in the planet's atmosphere. The disequilibrium of Earth's $O_2$ and $CH_4$ has been considered for decades as a template for life's expression (40). Methane is speculated to have been present in abundances high enough to warm Earth against the faint young Sun (41). However, the $O_2$ and $CH_4$ pair is unlikely to be detectable by JWST due to challenges of detecting $O_2$ (spectra feature at too short of a wavelength) and $O_3$, which is a proxy for $O_2$ (low instrument sensitivity at longer wavelengths). Further opposition to atmospheric chemical disequilibrium as a sign of life is the realization that such disequilibrium is a feature of most planets though the degree of chemical disequilibrium is variable. Lastly, in some cases, chemical disequilibrium may indicate the absence of life, as chemotrophic life forms utilize and diminish thermodynamic disequilibria in their environment, suggesting that disequilibrium could imply no life is present to exploit it. For example, methanotrophic bacteria take up $CH_4$ and $O_2$ and release $CO_2$, thus reducing the $CH_4$ (or $O_2$) levels in their environment to negligible levels.

Ideally, the community would develop highly sophisticated computer models to assess and even rule out false positives, by tracing a wide variety of gases through extensive planet evolution and ongoing interior and atmosphere processes. The presence of trace gases depends on specific planetary conditions that involve a layer of profound complexity and significant uncertainty (e.g., (42)). A large number of unknown parameters accompany the vast array of relevant geophysical processes, including: volcanic outgassing from—and ingassing to—planetary mantles; mantle evolution; mantle convection; magmatic outgassing; atmospheric escape; crustal oxidation; continental and sea floor weathering; deep volatile cycling; ongoing chemical segregation between a planetary core, mantle, and atmosphere, as well as surface, atmospheric and cloud non-equilibrium photochemical processes (e.g., (23)). One effort uses a Monte Carlo approach to sample a wide range of unknown parameters and initial conditions to connect magma ocean crystallization to temperate geochemical cycling (43). The overall complexity is problematic and adds heavily to the model inference problem by increasing the number of free parameters nearly without bound—most parameters are unobservable for exoplanets (Section 3).

Plausible Production Rates: Source vs. Sinks

To evaluate a potential biosignature gas we must consider its source rate vs. sink rate. The dominant sink is photochemistry (Figure 5), which is the key controlling factor for gas accumulation in an atmosphere. This is because trace gases, including biosignatures, are rapidly degraded by photochemical reactions. Gases are either directly destroyed by the host star's incident UV radiation or readily react with other abundant photochemically-produced atmospheric species, such as OH, H and halogen radicals. A first assessment step is to calculate the gas abundance required to generate a detectable spectral feature, through forward modeling. This step naturally includes a distinguishability criterion, namely that the gas should have one or more spectral features that are distinctive compared to anticipated background gases. The second step is to evaluate the candidate biosignature gas production rates, necessary for its accumulation in an exoplanet's atmosphere, against photochemical destruction and other sinks. A final step, if feasible, is to estimate the corresponding biomass (44). If the necessary production rates are excessively high then it is improbable for such gases to accumulate to detectable levels, or it might indicate an unrealistically large biomass. This framework intends to be pragmatic "reality check". Admittedly, "plausible" production rate is qualitative, but we may for example, use a comparison to Earth's production rates.

We emphasize that detectable levels that come out of computer simulations are gases with abundances of typically 1 to 100 ppm, much higher than biological or other trace gases of 1 ppt or 1 ppb typically reached in Earth's atmosphere (e.g., (5)). What is important to recognize is a candidate biosignature gas must saturate any sink, not just photochemical; surface sinks as well, including water oceans. Photochemical destruction rates are driven by the UV radiation from the host star and directly set the required biological production rates (45). The need for a deep understanding of photochemistry for biosignature gas studies echoes the giant exoplanet atmosphere frontier: a merging of astronomy and a new subbranch of astrochemistry (Section 2).

| Gas | Distinctive Spectral Features (vs dominant gas) | Plausible Production Rate | No Known Significant False Positives (Context) | Comments |
|---|---|---|---|---|
| $O_2$ | ✓ | ✓ | X | |
| $CH_4$ | ✓ | ✓ | X | |
| $N_2O$ | ✓ | (✓) | (✓) | |
| $CH_3Cl$ | ✓ | ✓ | (✓) | |
| $CH_3Br$ | ✓ | ✓ | (✓) | |
| $CH_3OH$ | ✓ | X | ✓ | Highly water soluble |
| $CH_3SH$ | ✓ | ✓ | X | |
| DMS | ✓ | ✓ | X | |
| $PH_3$ | ✓ | (✓) | ✓ | Relatively reactive |
| $NH_3$ | ✓ | X | ✓ | Highly water soluble |
| Isoprene | ✓ | X | ✓ | Reactive |
| Carbonyls | X | X | X | Highly water soluble |
| HCN | ✓ | ✓ | X | |
| $NO_2$ | ✓ | ✓ | X | |
| $SF_6$ | ✓ | X | ✓ | Technosignature gas |
| $NF_3$ | ✓ | X | ✓ | Technosignature gas |
| CFCs | ✓ | (✓) | ✓ | Technosignature gas |
| PFCs | ✓ | (✓) | ✓ | Technosignature gas |

Table 1. A list of potential biosignature gases with subjective evaluation. Reasonable production rate is subjective and may be qualified as compared to Earth values or Earth's $O_2$ production rate. See SI for a nearly exhaustive, referenced list of biosignature gases proposed so far.

### 4.2 A Growing List of Biosignature Gases

Many gases beyond $O_2$ and $CH_4$ have been studied as potential biosignature gases (see (46) and references therein). We present a summary in Table 1, followed by a nearly complete list of candidate biosignature gases studied to date in the SI. A glance at Table 1 will show that there are no biosignature gases that fulfill all the criteria for robust statements about the presence of life. We omit non-gas biosignatures (e.g., hazes, the red edge, algal bloom color changes, bioluminescence) as well as anti-biosignatures and false negatives, since robust assessment of these is infeasible in the JWST era (Section 3).

### 4.3 JWST's First Tentatively Claimed Biosignature Gas

We are off to a problematic start for biosignature gases with JWST, with the potential claim of dimethyl sulfide ($(CH_3)_2S$; DMS) in the atmosphere of the sub Neptune K2-18b (16). In this scenario K2-18b is a presumed "Hycean World" with a $H_2$-dominated atmosphere above a water ocean. The DMS detection or non detection depends on treatment of an instrumental offset which may cause a slight discontinuity in the data. Furthermore, the tentative 2.4$\sigma$ detection refers to the preference of a reference model consisting of 11 chemical species including 5 potential biosignature gases, including DMS, relative to a model with 10 chemical species of which 4 are potential biosignature gases not including DMS. That is, what is called a tentative detection hinges on a specific offset treatment and presumes that a model with 5 potential biosignatures is an appropriate reference. Regardless of offset treatment and parameter degeneracies, at face value, there is no robust statistical significance of the DMS spectral feature in the data. Instead, the detection relies on the retrieval process itself (See Section 3).

The concept of DMS in K2-18b is further explored by (47) with atmosphere computer simulations for DMS and other sulfur biosignature gases across a wide range of biological fluxes and stellar UV environments. For a mixture of DMS and other sulfur gases to reach JWST detectable levels (i.e. above ~ppm levels detectable in five transits) in K2-18b, the required volatile sulfur-compound biological flux is ~20 times higher than that of DMS' on modern Earth, as balanced against photochemical sinks.

The example of the tentative detection of DMS in K2-18b's atmosphere is the exoplanet community's first encounter with a biosignature gas prospect—a claim that fails all three Key Criteria above. The detection is not robust; the signal is not statistically significant (~3$\sigma$) and is sensitive to an instrumental offset. The attribution of the signal may be incorrect, as other sulfur gases share spectral features (48). Interpreting DMS as a biosignature must be confronted with potential abiotic formation scenarios (Table S2). The DMS attribution challenges echo lessons from the initial K2-18b $H_2O$ detection (24, 26) later argued to be (49) and confirmed as $CH_4$ (16). While higher SNR, higher spectral resolution data with broader wavelength coverage that encompass several DMS spectral features may help resolve the first two criteria, the third—interpretation—may remain unresolved due to unknown atmospheric contexts and processes.

For other suggested yet contentious biosignature gas reports, $PH_3$ on Venus, $CH_4$ on Mars, and HCN as a prebiotic gas on GJ 1132 b, see the SI.

## 5. Reflecting on the Future of Biosignature Detection

We conclude with the sobering realization that with JWST we may never be able to ***definitively*** claim the discovery of a biosignature gas in an exoplanet atmosphere. This realization is largely motivated by the challenge of the interpretation of false positives amidst the unknown planetary environments. But it is also exacerbated by the limited number of targets (Section 2), the likelihood that the spectral feature signals will be weak for small planet atmospheres (Section 2), and the limitations of the inference methods required to wrestle sparse data of a spatially and vertically unresolved planet into useful constraints on underlying planet atmosphere gas abundances (Section 3).

To confidently achieve the goal of identifying exoplanet biosignature gas candidates we should:

- Solve the effect of contamination by stellar magnetic activity;
- Discover long-period sub Neptune targets that are colder than our current crop and use atmosphere observations to infer if any may harbor a temperate water ocean;
- Refine our atmospheric retrieval methods;
- Build our list of plausible biosignature gases readily accessible by JWST, including the gases required for context.

On the theory side, we can develop a legitimate biosignature gas assessment pathway requiring a “Deep Planet Simulator”. We may strive to develop comprehensive models for exoplanet interiors, akin to the elaborate Illustris cosmological simulation of galaxy formation (50). However, the cosmologists have a “unified theory” to test and refine, whereas the exoplanetary community is far from any unified theory of planetary habitability. By modeling from deep cores through dynamic mantles to complex atmospheres, we could seek to understand the conditions that might support identification of biosignature gases over false positives. However, unlike for galaxy formation and evolution models which were bolstered by increasing spatial resolution via new telescopes over time, there is no way to validate exoplanet interior computer models using remote sensing of spatially averaged atmospheres. As components of these models evolve and integrate experimental data, their credibility in identifying potential biosignatures will be a focal point of rigorous debate. Lab experiments need to be expanded on, as well as further synergies with the Solar System community.

Beyond JWST, we can pursue the discovery of an Earth twin around a Sun-like star using future high-contrast starlight-suppression direct imaging such as space-based mission concepts like NASA’s Habitable Worlds Observatory, Starshade (51), and the Large Interferometer for Exoplanets (LIFE) (52), or even ground-based observatories (53). A familiar scenario, with an Earth-like atmospheric mix of $O_2$ and water (as opposed to a super-Earth around an M dwarf star) might yield more definitive answers, but likely will still face the wide unknowns of planetary context.

We can investigate technosignature gases—artificially produced volatiles, either intentional or accidental, by advanced civilization (54). For example, fluorine is largely avoided by life on Earth but widely used in human-made products. While technosignatures might overlap with biosignature gases if alien life uses fluorine, fluorine-based technosignatures have few, if any, false positives (Table S2).

Ultimately, we seek fundamentally new technological approaches for exoplanet observations. Real breakthroughs will come from audacious projects like the Solar Gravitational Lens telescope, positioned 500 AU from Earth to exploit the Sun as a gravitational lens (55), or from the Starshot initiative, which envisions sending thousands of space chips with solar sails to pass by a planet and capture brief, yet potentially revealing, snapshots to send back to Earth.

The exoplanet community has come a long way in 30 years, establishing that exoplanets are common, and that rocky planets exist, including some that may have surface liquid water. We expect to find biosignature gas candidates, even if we cannot guarantee they are signs of life—an uncertainty we will have to live with for now. In the years to come, JWST will remain the flagship of this era of discovery, and will be remembered as the first telescope that set the first concrete steps toward answering the question: Are we alone?

**Acknowledgements**
We thank Julien de Wit and Renyu Hu for useful discussions.

# Supplementary Information for

Prospects for Detecting Signs of Life on Exoplanets in the JWST Era.

Sara Seager[1,2,3,*], Luis Welbanks[4], Lucas Ellerbroek[5], William Bains[6], Janusz J. Petkowski[7,8]

[1] Department of Earth, Atmospheric and Planetary Sciences, Massachusetts Institute of Technology, Cambridge, MA, USA
[2] Department of Physics, Massachusetts Institute of Technology, Cambridge, MA, USA
[3] Department of Aeronautical and Astronautical Engineering, Massachusetts Institute of Technology, Cambridge, MA, USA
[4] School of Earth and Space Exploration, Arizona State University, Tempe, AZ, USA
[5] Department of Astrophysics / IMAPP, Radboud University, PO Box 9010, 6500 GL, Nijmegen, The Netherlands
[6] School of Physics and Astronomy, Cardiff University, 4 The Parade, Cardiff CF24 3AA, UK
[7] JJ Scientific, Warsaw, Mazowieckie, Poland
[8] Faculty of Environmental Engineering, Wroclaw University of Science and Technology, 50-370 Wroclaw, Poland

*Correspondence: Sara Seager
Email: seager@mit.edu

**This PDF file includes:**

Supplementary text
Tables S1 to S2
SI References

**Supplementary Information Text**

**1. Biosignature Gas Candidate Detections Preceding the JWST Era.**

There are few reports of biosignature gases detections that precede the coming era of the JWST exoplanet atmosphere studies – the reports of phosphine ($PH_3$) on Venus (1), methane ($CH_4$) on Mars (2) and hydrogen cyanide (HCN) and $CH_4$, as prebiotic molecules, on GJ 1132b (3). All three cases exemplify the difficulty of such observations, and the inherent uncertainty of data interpretation that will challenge the JWST search for genuine signs of life. We summarize these examples briefly below.

The detection of an atmosphere on a rocky Earth-size exoplanet, GJ 1132b has been reported in 2021. The Hubble WFC3 infrared atmospheric transmission spectrum has been claimed to show spectral signatures of aerosol scattering, HCN, and $CH_4$ gas in a low mean molecular weight atmosphere (3). Other groups that reanalyzed the same Hubble data find a featureless spectrum, with no evidence of an atmosphere (4, 5). The initial observations of GJ 1132b have also recently been followed up with JWST. The thermal emission measurements of GJ 1132b with the Mid-Infrared Instrument Low-Resolution Spectrometer (MIRI), in 5–12 μm, support the conclusion that GJ 1132b likely does not have a significant atmosphere (6).

The search for biosignature gases in our on solar system is also controversial. The detection of $CH_4$ on Mars was first reported in 2004 from both a ground-based telescope (2) and by the Mars Express Orbiter (7). Both found a weak signal with small values, 10 ± 3 ppb and 10 ± 5 ppb respectively. The initial observations were followed by in situ detection by the TLS component of the SAM instrument on the Mars Curiosity Rover (e.g. (8, 9)).

There is a continuing discussion on the existence of $CH_4$ on Mars. The Trace Gas Orbiter (TGO) does not observe methane and set a detection limit of 0.05 ppb despite hundreds of observations (10). Criticism that the ground-based detections by (11) were simply a detection of the $^{13}CH_4$ isotope in Earth's atmosphere (12) were later refuted (13). The in situ TLS/SAM findings of seasonal methane (9) come from an instrument designed to avoid contamination (14), but have been argued to be contamination (15) or noise (16).

Methane is not expected to be present on Mars because it is a reduced gas and the Martian atmosphere and surface are oxidizing environments. The presence of $CH_4$ would indicate an unusual geochemistry or leave the possibility for the presence of life. The relatively short ~300 year photochemical lifetime of $CH_4$ means that it must have a current-day (or very recent) source. This source, if present, is unknown. For a summary of the Martian $CH_4$ debate see (14, 17).

Phosphine ($PH_3$) gas in the Venus atmosphere was reported at a few ppb levels in 2020 (1) and followed up by evidence of $PH_3$ from a data reanalysis of NASA's 1978 Pioneer Venus Probe's mass spectrometer (18). While additional astronomical observations have supported the presence of $PH_3$ (19), the majority do not detect the signal at all (20–22), instead ascribing the result to statistical or processing errors (23–25). Some others do recover the signal and propose gas attribution to $SO_2$ rather than $PH_3$ (26, 27).

On Venus and Earth $PH_3$ should not be present due to the oxidizing environment and short $PH_3$ photochemical lifetime. On Earth, $PH_3$ only has biological or industrial sources (e.g. (28)). Work has largely demonstrated that known abiotic chemistry—lightning, volcanoes, meteoritic delivery—in the Venus environment does not produce the required amounts of $PH_3$ to explain claimed levels (e.g. (29, 30)), though future research may find viable pathways due to unknown $PH_3$ chemistry (Table S2). For a recent review of observational and interpretation claims and refutations, see (31).

That the detection of methane on Mars is still not accepted 20 years after its first report, despite detections from Earth, Mars orbit, and Mars highlights the challenges in addressing trace gases for questions [1] and [2], let alone the more difficult question [3]. Here questions are defined in the main manuscript Section 4. Similarly, the ongoing debate on the presence of $PH_3$ on Venus shows that clear answers to questions [1] and [2] likely will not be resolved without a dedicated space mission with an in-situ measurement capability. The answer to question [3] also remains inconclusive (Table S2).

Although both Mars and Venus are challenging environments for which to imagine life's existence—Mars life needing to be confined subsurface and Venus' to temperate cloud layers composed of sulfuric acid—both are neighboring planets which we know far more about than we will know about any exoplanet. One may have thought if there ever was a place where it would be possible to confirm a biosignature, it would be a planet next door, especially as we can and have sent in situ probes to physically sample the environment. Sampling an exoplanet remains in the realm of science fiction.

**Table S1.** Selected biosignature gases and their JWST observational prospects. The table shows selected JWST observational prospects focusing on the diverse stellar and planetary scenarios. We only show published computer-simulated predictions, not the proposed, or approved, JWST observational campaigns.

| **Biosignature Gas** | **Stellar Type** | **Planet Type** | **Atm. Type** | **JWST Detectable Flux or Abundance / at λ (μm)** | **JWST Observation Time** |
|---|---|---|---|---|---|
| $O_2/O_3$ | Trappist-1 (M8V) | Rocky Earth size (Trappist-1e) | Archean Earth | Not detectable (32–37) | N/A |
| $CH_4/CO_2$ pair | Trappist-1 (M8V) | Rocky Earth size (Trappist-1e) | Archean Earth | Archean-Earth-like $CH_4$ levels | 10 transits (35, 38, 39) |
| $CH_4/CO_2$ pair | LHS 1140 (M4.5) | Sub Neptune (LHS 1140 b) | $H_2$ | ? / NIRSpec [1-5 μm] | 1 transit (40) |
| $N_2O$ | Trappist-1 (M8V) | Rocky Earth size (Trappist-1e) | $N_2$-$O_2$ | 10–100 Tmol $yr^{-1}$ / 2.9 μm | detectable "within [JWST] mission lifetime" (41) |
| $N_2O$ | LHS 1140 (M4.5) | Sub Neptune (LHS 1140 b) | $H_2$ | ? / NIRSpec [1-5 μm] | 10–50 transits (40–200 hours) (40) |
| $CH_3Cl/CH_3Br$ pair | Trappist-1 (M8V) | Rocky Earth-size (Trappist-1e) | $N_2$-$O_2$ | $10^{12}$ molecules $cm^{-2}$ $s^{-1}$ / MIRI-LRS [5-14 μm] | 10 transits (42) |
| $CH_3Cl$ | LHS 1140 (M4.5) | Sub Neptune (LHS 1140 b) | $H_2$ | ? / NIRSpec [1-5 μm] | 10–50 transits (40–200 hours) (40) |
| $CH_3OH$ | M5V (based on GJ 876) | Rocky super Earth (1.5 $R_{Earth}$, 5 $M_{Earth}$) | $H_2$ | 10 ppm / G395M [2.87-5.14 μm] and MIRI-LRS [5-14 μm] | 20 transits (43) |
| $CH_3SH$ | Trappist-1 (M8V) | Rocky Earth size (Trappist-1e) | Archean Earth | 30 x Earth flux rate / NIRSpec [1-5 μm] | 50 transits (35) |
| DMS | K2-18 (M2.8) | Sub Neptune (K2-18b) | $H_2$ | 20 x Earth flux rate / MIRI-LRS [5-14 μm] | 5 transits (44) |
| DMS | Trappist-1 (M8V) | Rocky Earth size (Trappist-1e) | Archean Earth | 30 x Earth flux rate / NIRSpec [1-5 μm] | 94 transits (35) |
| $PH_3$ | Active M-dwarf | Rocky super Earth (1.75 $R_{Earth}$, 10 $M_{Earth}$) | $H_2$ | 4 ppm / 7.8-11.5 μm in emission | 52 hours (28) |
| $PH_3$ | Active M-dwarf | Rocky super Earth (1.75 $R_{Earth}$, 10 $M_{Earth}$) | $CO_2$ | 310 ppm / 7.8-11.5 μm in emission | 48 hours (28) |
| $PH_3$ | LHS 1140 (M4.5) | Sub Neptune (LHS 1140 b) | $H_2$ | $10^{10}$ molecules $cm^{-2}$ $s^{-1}$ / NIRSpec [1-5 μm] | 10–50 transits (40–200 hours) (40) |
| $NH_3$ | M5V | Rocky super Earth (1.75 $R_{Earth}$, 10 $M_{Earth}$) | $H_2$ | 5 ppm NIRSpec [1-5 μm] and MIRI-LRS [5-14 μm] | 80 transits (45) |
| $NH_3$ | M4V (based on GJ 1132) | Rocky super Earth (1.75 $R_{Earth}$, 10 $M_{Earth}$) | $H_2$ | 14 ppm / MIRI-LRS [5-14 μm] | 3 transits (46) |
| $NH_3$ | LHS 1140 (M4.5) | Sub Neptune (LHS 1140 b) | $H_2$ | $8.4x10^{10}$ molecules $cm^{-2}$ $s^{-1}$ / NIRSpec [1-5 μm] | 10–50 transits (40–200 hours) (40) |
| Isoprene | M5V | Rocky super Earth (1.75 $R_{Earth}$, 10 $M_{Earth}$) | $H_2$ | $3x10^{13}$ molecules $cm^{-2}$ $s^{-1}$ / NIRSpec [1-5 μm] and MIRI | 20 transits (47) |
| Formaldehyde ($CH_2O$) and other carbonyls | M4V (based on GJ 876) | Rocky super Earth (1.75 $R_{Earth}$, 10 $M_{Earth}$) | $H_2$ | Not detectable (48) | N/A |
| Formaldehyde ($CH_2O$) and other carbonyls | M4V (based on GJ 1132) | Rocky super Earth (1.75 $R_{Earth}$, 10 $M_{Earth}$) | $H_2$ | 3.1 ppm / MIRI-LRS [5-14 μm] | 3 transits (46) |
| HCN | M4V (based on GJ 1132) | Rocky super Earth (1.75 $R_{Earth}$, 10 $M_{Earth}$) | $H_2$ | 1.7 ppm / G395M [2.87-5.14 μm] | 3 transits (46) |
| $NO_2$ | M dwarf | Rocky Earth size | $N_2$ | 20 x present Earth's atm. abundance | 500 hours (49) |
| $SF_6$ | M5V (based on GJ 876) | Rocky super Earth (1.5 $R_{Earth}$, 5 $M_{Earth}$) | $H_2$ | 1 ppm / 9-12 μm | Tens of transits (50) |
| $SF_6$ | Trappist-1 (M8V) | Rocky Earth size (Trappist-1f) | $N_2$-$O_2$ | 100 ppm / MIRI-LRS [5-14 μm] | 19 transits (51) |
| $NF_3$ | M5V (based on GJ 876) | Rocky super Earth (1.5 $R_{Earth}$, 5 $M_{Earth}$) | $H_2$ | 1 ppm / 9-12 μm | Tens of transits (50) |
| $NF_3$ | Trappist-1 (M8V) | Rocky Earth size (Trappist-1f) | $N_2$-$O_2$ | 100 ppm / MIRI-LRS [5-14 μm] | 16 transits (51) |
| CFCs | White dwarf | Rocky Earth size | $N_2$ | 10 x present Earth's atm. abundance | 30-40 hours (52) |
| CFCs | M dwarf | Rocky Earth size | $N_2$ | >1 ppb | 100-300 hours (53) |
| PFCs | Trappist-1 (M8V) | Rocky Earth size (Trappist-1f) | $N_2$-$O_2$ | 100 ppm / MIRI-LRS [5-14 μm] | 5 transits (51) |

**Table S2.** False positive assessment of biosignature gases.

| **Biosig. Gas** | **Proposed Abiotic Source** |
|---|---|
| $O_2$/$O_3$ | Massive $O_2$ atmospheres (with partial pressures $pO_2 \geq 10$ bar) are postulated to exist via abiotic accumulation on M-dwarf-hosted exoplanets. The main proposed mechanism is XUV-driven photodissociation and escape during the extended pre-main sequence runaway greenhouse phase (54). Terrestrial planets orbiting any star type may develop $O_2$-dominated atmospheres as a result of $H_2O$ photolysis due to inefficient cold trap (55). Massive $O_2$ atmospheres are thought to only form abiotically, and can be identified by the presence of $O_2$-$O_2$ dimers (56, 57). Abiotic $O_2$ and $O_3$ can also form through planetary photochemistry of $CO_2$ (58, 59). Conditions for $O_2$ false positives on habitable zone planets around Sun-like stars have also been explored (60). |
| $CH_4$ | Methane can be produced geologically, e.g., in the serpentinization reaction but large amounts of $CH_4$ are unlikely to have volcanic origin (61). $CH_4$ can also be produced by impacts (62, 63). Its potential as a biosignature highly depends on the stellar and planetary context (64). |
| $N_2O$ | Significant abiotic sources of $N_2O$ are limited. Potential abiotic sources like reduction of nitric oxide (NO) by ferrous iron (chemidenitrification), lightning, volcanic activity, or photochemical reduction of NO can be ruled out or identified by relevant stellar and planetary context (41). |
| $CH_3Cl$ | Meteoritic infall and volcanism could be the source of small amounts of abiotic $CH_3Cl$ (42). |
| $CH_3Br$ | Meteoritic infall and volcanism could be the source of small amounts of abiotic $CH_3Br$ (42). |
| $CH_3OH$ | There are no known significant abiotic $CH_3OH$ sources on terrestrial planets (43). |
| $CH_3SH$ | Geological and hydrothermal sources are unlikely to generate $CH_3SH$ in significant amounts (65). Laboratory photochemical experiments demonstrate the abiotic production of organosulfur gases (66). |
| DMS | Geological and hydrothermal sources are unlikely to generate DMS in significant amounts (65). Laboratory photochemical experiments demonstrate the abiotic production of organosulfur gases (66). DMS may also be produced abiotically in cometary matter (67) and the interstellar medium (68). |
| $PH_3$ | No known significant abiotic source on rocky exoplanets (28). In the context of Venus: no significant conventional abiotic sources known (29, 31). Volcanoes are an inefficient source of $PH_3$ (30). Other potential abiotic sources like reduction with Fe-rich minerals (69) or synthesis over acidic dust in Venus' atmosphere await confirmation (70). |
| $NH_3$ | No known significant abiotic source on rocky exoplanets (45). Volcanic, hydrothermal, and other processes (e.g. nitrogen photoreduction on $TiO_2$ containing sands) are too inefficient to result in a detectable $NH_3$ (45). |
| Isoprene | Isoprene ($C_5H_8$) has no known false positive sources (47). |
| Carbonyls | Formation of formaldehyde ($CH_2O$) via photochemical processes is potentially possible (46, 71) although its accumulation in the atmosphere to significant amounts is unlikely (48). |
| HCN | Hydrogen cyanide (HCN) can be produced via UV photochemistry (46, 72–75), surface hydrothermal systems (76), impacts (77, 78), and lightning (79). |
| $NO_2$ | Produced via lightning from atmospheric $N_2$ and $CO_2$ (65). |
| $SF_6$ | Sulfur hexafluoride ($SF_6$) has very few false positives. Trace amounts of $SF_6$ are associated with volcanic activity (80–82). $SF_6$ can be released from fluorite minerals (83). H-depleted planets could potentially produce abiotic $SF_6$ in greater amounts than on Earth (50). |
| $NF_3$ | Nitrogen trifluoride ($NF_3$) has no known significant false positive sources. $NF_3$ is not known to be released from any fluorine-containing minerals apart from a single exception of purple fluorite (WF) (83). |
| CFCs | Various CFCs can be produced in trace amounts by volcanoes (e.g. (84, 85)) or can be detected in trace amounts in fluorite minerals (83). |
| PFCs | Abiotic $CF_4$ is produced geologically in trace amounts (e.g. (81–83)). |

## Supplementary References